\documentclass[showpacs,amsmath,amssymb,pre,aps,superscriptaddress,twocolumn]{revtex4-1}
\usepackage[tight,TABTOPCAP]{subfigure}
\usepackage[usenames, dvipsnames]{color}
\usepackage{natbib}
\usepackage[utf8]{inputenc}
\usepackage{textcomp}
\usepackage[english]{babel}
\usepackage{amsmath}
\usepackage{amsfonts}
\usepackage{amssymb}
\usepackage{makeidx}
\usepackage{graphicx}
\usepackage{lipsum}
\usepackage{tikz}
\usepackage{float}
\usetikzlibrary{arrows}
\usetikzlibrary{decorations.pathreplacing,decorations.markings}
\usetikzlibrary{mindmap, trees}
\usetikzlibrary{shapes.geometric}
\usepackage{pgfplots}
\pgfplotsset{compat=1.16}
\usetikzlibrary{3d}
\usepackage[left=2cm,right=2cm,top=2cm,bottom=2cm]{geometry}
\usepackage{hyperref}

\begin{document}

\title{Finite-size excess-entropy scaling for simple liquids}
\author{Mauricio Sevilla}
\affiliation{Max Planck Institute for Polymer Research, Ackermannweg 10, 55128, Mainz}
\author{Atreyee Banerjee}
\affiliation{Max Planck Institute for Polymer Research, Ackermannweg 10, 55128, Mainz}
\author{Robinson Cortes-Huerto}
\email{corteshu@mpip-mainz.mpg.de}
\affiliation{Max Planck Institute for Polymer Research, Ackermannweg 10, 55128, Mainz}
\date{\today}
\begin{abstract}
We introduce and validate a finite-size two-body excess entropy integral equation. By using analytical arguments and computer simulations of prototypical simple liquids, we show that the excess entropy $s_2$ exhibits a finite-size scaling with the inverse of the linear size of the simulation box. Since the self-diffusivity coefficient $D^*$ displays a similar finite-size effect, we show that the scaling entropy relation $D^*=A\exp(\alpha s_2)$ also depends on the simulation box size. By extrapolating to the thermodynamic limit, we report values for the coefficients $A$ and $\alpha$ that agree well with values available in the literature. Finally, we find a power law relation between the scaling coefficients for $D^*$ and $s_2$, suggesting a constant viscosity to entropy ratio.  
\end{abstract}
\maketitle
\section{Introduction}
\begin{figure}[h!]
	\centering
	\includegraphics[width=0.48\textwidth]{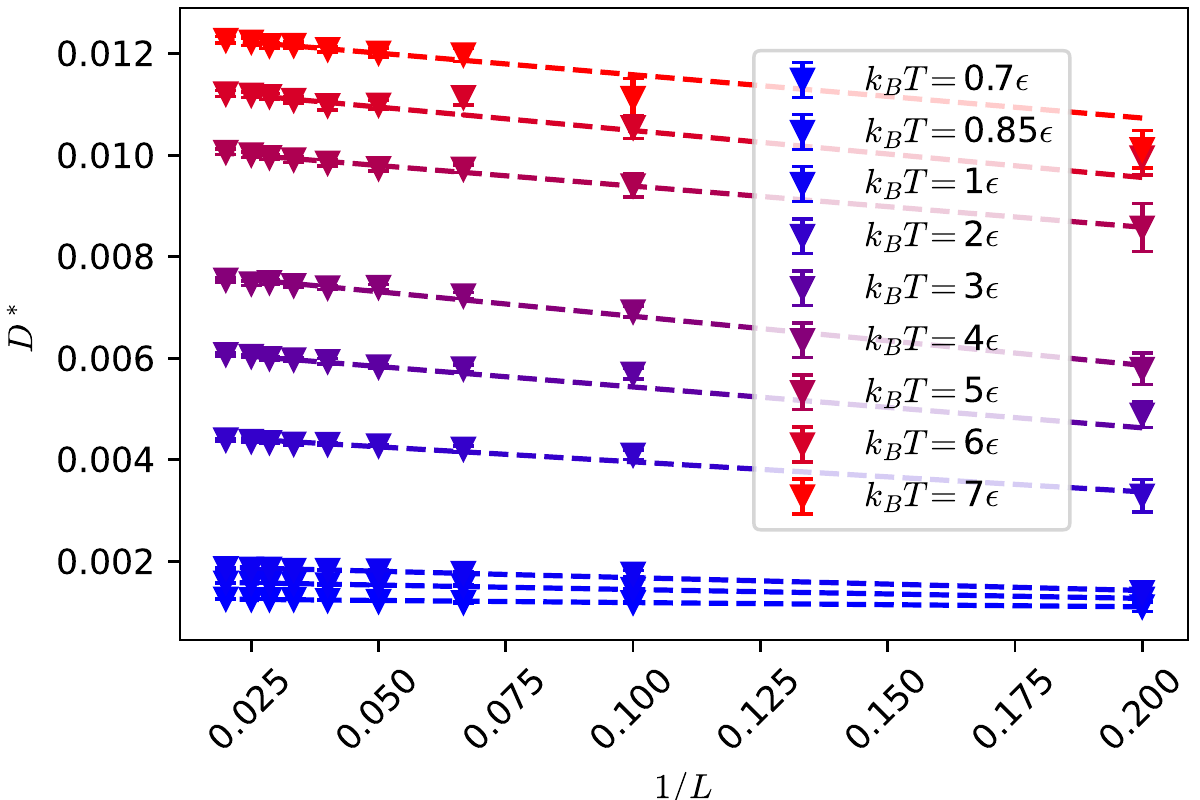}
	\caption{
		Reduced self-diffusion coefficient $D^*$ as a function of the inverse of the box linear size $1/L$ for a Lennard-Jones liquid with density $\rho\sigma_{\mathrm{LJ}}^3=0.864$ in the range of temperatures $k_{\rm B}T=[0.7\epsilon, 7 \epsilon]$. 	
	} \label{fig:Figure_01}
\end{figure}

Excess entropy ($s_{\rm exc}$), the difference between the entropy of a system and its ideal gas counterpart at the same temperature and density, is connected to the dynamical properties of simple liquids (See Ref.~\cite{dyre2018perspective} for a recent review). This observation was first reported by Rosenfeld~\cite{Rosenfeld1977,*Rosenfeld1999quasi}, who showed that, for simple model liquids,  reduced transport properties such as diffusivity, viscosity and thermal conductivity scale with the excess entropy as
\begin{equation}
X^* = A\exp\left(\alpha s_{\rm exc}\right)\, ,
\label{eq:trans_sexc}
\end{equation}
with $X^*$ a dimensionless transport property and $A$ and $\alpha$ parameters, independent of the thermodynamic state, determined by the interparticle potential. \\
Following similar physical arguments and assuming that the major contribution to $s_{\rm exc}$ comes from two-body terms, Dzugutov proposed a similar scaling relation between self-diffusivity and a two-body approximation to the excess entropy $s_2$, namely~\cite{Dzugutov1996}
\begin{equation}
D^* = A\exp(\alpha s_2)\, ,
\label{eq:Dzugutov}
\end{equation}
with $D^*=\frac{D}{\Gamma \sigma_{r}^2}$ where $D$ is the self-diffusion coefficient, $\sigma_r$ measures the linear size of the particles and $\Gamma=4\sigma^2g(\sigma_r)\rho\sqrt{\frac{\pi k_{\rm B}T}{m}}$ the collision frequency given by the Enskog theory~\cite{chapman1990} where $g(\sigma_r)$ is the value of the radial distribution function at a distance $\sigma_r$. In this case, a large variety of simple liquids satisfy Eq.~\eqref{eq:Dzugutov} with the \emph{universal} choice of parameters $A=0.049$ and $\alpha=1$~\cite{Dzugutov1996}.\\ 
This excess entropy scaling has been widely validated for a large variety of simple~\cite{PhysRevLett.85.594,doi:10.1063/1.1516594,PhysRevE.68.031204,Zhu2005,Jeff2011,Atreyee_JChemSci2017,widom2019first} and molecular liquids\cite{Goel_etal_JCP2008,Malvaldi_Chiappe_JCP2010,Chopra_etal_JCP2010,Galliero_etal_JCP2011}, including specially water~\cite{doi:10.1080/00268970802378662,Agarwal_etal_JPCB2010,chopra2010use,Agarwal_JPCB2011}. We also highlight that experimental studies have tested entropy scaling in somewhat challenging scenarios~\cite{Tong2013,spieckermann2022structure}, and the fact that Rosenfeld and Dzugutov relations are empirical but have been justified on theoretical grounds~\cite{Alok_etal_PRL2001,seki2015relationship}. Furthermore, the structure--dynamics connection in Eq.~\eqref{eq:Dzugutov} has been proposed as a tool to investigate the relation between dynamical properties of computational models at different resolutions,~\cite{Jeff2012}, which is now routinely considered in the context of coarse-grained models~\cite{rondina2020predicting,jin2022understanding}.  \\
Transport properties exhibit implicit size effects due to the finite size of the simulation box and the use of periodic boundary conditions (PBC)~\cite{Duenweg1993,Pascal_PRB2016,Zaoui_etal_PRB2016}. In the particular case of the reduced self-diffusion coefficient $D^*$, given a cubic simulation box of linear size $L$, $D^{*}\equiv D^{*}(L)$ takes the form~\cite{Duenweg1993,Hummer2004,Kikugawa_etal_JCP2015I, Kikugawa_etal_JCP2015II, Rotenberg2015} (See Figure \ref{fig:Figure_01})
\begin{equation}\label{eq:DvsL}
D^*(L) = D^{*\infty} - \frac{\delta}{L}\, ,
\end{equation}
with $\delta=\frac{k_{\rm B} T \zeta}{6\pi \eta \Gamma \sigma_r^2}$ with $\zeta\approx 2.837297$ and $\eta$ the system's viscosity.  In the thermodynamic limit (TL), namely, in the limit $L\to \infty$, the self-diffusion coefficient takes the value $D^{*\infty}$.\\
Given the finite-size scaling of $D^*$, we expect that Eq.~\eqref{eq:Dzugutov} also depends on the size of the simulation box. Recent computational studies investigating entropy scaling for liquid water using ab initio molecular dynamics simulations~\cite{Herrero_etal_PNAS2022} emphasise the relevance of this remark. In this case,  the systems under consideration are rather small, and finite-size effects become increasingly important.\\
In this paper, we investigate the finite-size scaling of Eq.~\eqref{eq:Dzugutov} by focusing on  implicit and explicit finite-size effects present on the two-body excess entropy $s_2$. We find that $s_2$ obeys a finite-size scaling relation similar to $D^*$, which implies that the \emph{universal} parameters $A$ and $\alpha$ in Eq.~\eqref{eq:Dzugutov} also depends on the size of the simulation box. Finally, and perhaps more interestingly, our results indicate that a power law relates the finite-size scaling coefficients of $D^*$ and $s_2$, suggesting a constant viscosity/entropy ratio~\cite{Kovtun_etal_PRL2005, Angiella_etal_PhysLettA2009, Faussurier_etal_HEDP2014, Hohm_ChemPhys2014}.\\
The paper is organised as follows:  In Section  \ref{sec:C} we present the model and computational details.  We show that $s_2$ is ensemble invariant and that the only relevant finite-size effect comes from using finite integration domains in Section \ref{sec:EI}. In Section \ref{sec:fvs2}, we introduce and validate a finite-size version of $s_2$. We then present the finite-size scaling of the Dzugutov relation (Eq.~\eqref{eq:Dzugutov}) in Section \ref{sec:fsDs2}. Finally, we conclude and provide our outlook in Section \ref{sec:Con}.
\section{Computational details}\label{sec:C}
We investigate the excess entropy scaling for liquids whose potential energy is described by a 12--6
Lennard--Jones potential truncated, with cutoff radius $r_{c}/\sigma_{\mathrm{LJ}}=2.5$, and shifted. The parameters $\epsilon$, $\sigma_{\mathrm{LJ}}$ and $m$, define the energy, length and mass units, respectively. All the results are expressed in 
LJ units with time $\sigma_{\mathrm{LJ}}(m/\epsilon)^{1/2}$, 
temperature $\epsilon/k_{\rm B}$ and pressure $\epsilon/\sigma_{\mathrm{LJ}}^{3}$. In the following, we identify $\sigma_r$ of Eq.~\eqref{eq:DvsL} with $\sigma_{\mathrm{LJ}}$. We consider cubic simulation boxes with linear sizes in the interval $L/\sigma_{\mathrm{LJ}} = [5,50]$, with fixed density $\rho\sigma_{\mathrm{LJ}}^{3}=0.864$. The systems are equilibrated at temperatures in the interval $k_{\rm B}T=[0.7\epsilon, 7.0\epsilon]$, enforced with a Langevin thermostat with damping coefficient $\gamma (\sigma (m/\epsilon)^{1/2}) = 1.0$. We equilibrate the samples for $10 \times 10^{6}$ molecular dynamics (MD) steps using a time step of $\delta t/(\sigma_{\mathrm{LJ}} (m/\epsilon)^{1/2}) = 10^{-3}$, followed by additional $10 \times 10^{6}$ MD steps on the NVE ensemble to verify that the temperature does not deviate substantially from the target value. Production runs span $10 \times 10^{6}$ MD steps.
All the simulations have been performed with the LAMMPS simulation package \cite{LAMMPS}.
\section{Implicit and explicit finite-size effects}\label{sec:EI}
In this section, we identify which finite-size effects are expected to affect the calculation of the excess entropy. We start with the definition of excess entropy for an $N$--particle system with respect to the ideal gas:
\begin{equation}
s_{\rm exc} = \frac{S-S_{\rm IG}}{N k_{\rm B}} = \frac{S_{2}+S_{3}+\cdots}{N k_{\rm B}}\, ,
\end{equation}
with $k_{\rm B}$ the Boltzmann constant. In the following, we focus on two-body contributions, which mostly amount to 80--90$\%$ of the overall value of the excess entropy for simple liquids.~\cite{borzsak1992convergence,banerjee2014role} In particular, we have~\cite{Raveche_JCP55_2242_1971,Mountain_Raveche_JCP55_2250_1971}
\begin{equation}\label{eq:s2}
s_2 = -\frac{\rho}{2 V}\int_V \int_V d\mathbf{r}_1\, d\mathbf{r}_2\, \left[ g(\mathbf{r})\ln g(\mathbf{r}) - (g(\mathbf{r}) -1)  \right]\, ,
\end{equation}
with $s_2=\frac{S_{2}}{N k_{\rm B}}$ the two-body excess entropy per particle.  By taking the thermodynamic limit and assuming that the liquid is homogeneous and isotropic, we obtain the familiar expression
\begin{equation}\label{eq:s2_TL}
s_2^{\infty} = -2\pi \rho\int_{0}^{\infty} dr\, r^{2}\left[ g(r)\ln g(r) - (g(r) -1)  \right]\, .
\end{equation}
When performing molecular dynamics simulations, we usually consider systems with a finite number of particles, typically not large enough to reach the thermodynamic limit. Furthermore, when evaluating the double integral in Eq.~\eqref{eq:s2} we need to consider that the volume $V$ is finite. For such a reason, and following the strategy used to compute the compressibility equation \cite{Roman2008,CortesHuerto_Entropy2018} and the Kirkwood-Buff integrals \cite{schnell-etal-JPhysChemLett4-235-2013,CortesHuerto_Communication2016,Robin2018b} in computer simulations, we define a finite--size two--body excess entropy evaluated in a subvolume $V$ of a system with a total number of particles $N_0$ and the volume $V_0$
\begin{equation}\label{eq:s2_finite}
\begin{split}
s_2(V;N_0) = -\frac{\rho}{2V}\int_V \int_V & d\mathbf{r}_1\,  d\mathbf{r}_2\, \left[ g(\mathbf{r};N_0)\ln g(\mathbf{r};N_0)\right. \\
&\left. - (g(\mathbf{r};N_0) -1)  \right]\, ,
\end{split}
\end{equation}
with $g(\mathbf{r};N_0)$ the finite-size RDF. The asymptotic correction to the finite-size RDF, given by the difference in the thermodynamic ensemble, gives~\cite{Percus1961,Percus1961II,Salacuse-etal-PRE53-2382-1996,Roman-etal-JChemPhys107-4635-1997,Villamaina-Trizac-EurJPhys35-035011-2014,Roman-etal-AmJPhys67-1149-1999}
\begin{equation}
g(\mathbf{r};N_0)  = g(\mathbf{r}) -\frac{\chi_{T}^{\infty}}{N_0}
\end{equation}
with $\chi_{T}^{\infty}=\rho k_{\rm B} T\kappa_T$, and $\kappa_T$ being the isothermal compressibility in the thermodynamic limit.  We write the integrand in Eq.~\eqref{eq:s2_finite} as 
\begin{equation}
\begin{split}
&g(\mathbf{r};N_0)\ln g(\mathbf{r};N_0)\approx g(\mathbf{r})\ln g(\mathbf{r}) \\
&\quad \quad \quad - \frac{\chi_T^{\infty}}{N_0}(1 + \ln g(\mathbf{r}))\\
&g(\mathbf{r};N_0) - 1 = g(\mathbf{r}) - 1 - \frac{\chi_T^{\infty}}{N_0}\, ,
\end{split}
\end{equation}
where in the first line in the previous expression, we have neglected terms of the order $O\left(\frac{1}{N_0^2}\right)$.  The two contributions $\frac{\chi_T^{\infty}}{N_0}$ cancel out exactly. The contribution $\frac{\chi_T^{\infty}}{N_0}\ln g(\mathbf{r})$ can be neglected by assuming a large number of particles (there is no $V/V_0$ contribution, only $1/V_0$, hence, we can neglect it). This indicates that the two-body excess entropy is ensemble invariant, consistent with the result reported Ref.~\cite{Wallace1987,Baranyai1989}. We thus rewrite Eq.~\eqref{eq:s2_finite} as
\begin{equation}
\begin{split}
s_{2}(V) = -\frac{\rho}{2 V}\int_V \int_V d\mathbf{r}_1\, \ d\mathbf{r}_2\, &\left[ g(\mathbf{r})\ln g(\mathbf{r})\right.\\
&\left. - (g(\mathbf{r}) -1)  \right]\, .
\end{split}
\end{equation}
The volume $V$ is finite and embedded into the volume $V_0$.  The integration domains can be rearranged as $\int_V\int_V(\cdots) = \int_V\int_{V_{0}}(\cdots) - \int_V\int_{V_{0}-V}(\cdots) $. Using a similar argument as the one used to calculate the finite-size compressibility~\cite{Binder-etal-EPL6-585-1988} and Kirkwood-Buff integrals~\cite{schnell-etal-JPhysChemLett4-235-2013},  the term $\int_V\int_{V_{0}}(\cdots)$ gives $s_{2}^{\infty}$ and the term $\int_V\int_{V_{0}-V}(\cdots) $ scales as $1/L$ with $L=V^{1/3}$ the linear size of the cubic simulation box. Thus, 
\begin{equation}\label{eq:s2_1overL}
s_{2}(L) = s_{2}^{\infty} + \frac{\sigma}{L}\, ,
\end{equation}
with $\sigma$ a constant that depends on intensive thermodynamic quantities only.  In the following section, we introduce a method to compute $s_{2}(L)$ and verify its scaling behaviour with the linear size of the simulation box.\\
\begin{figure}[h!]
	\centering
	\includegraphics[width=0.48\textwidth]{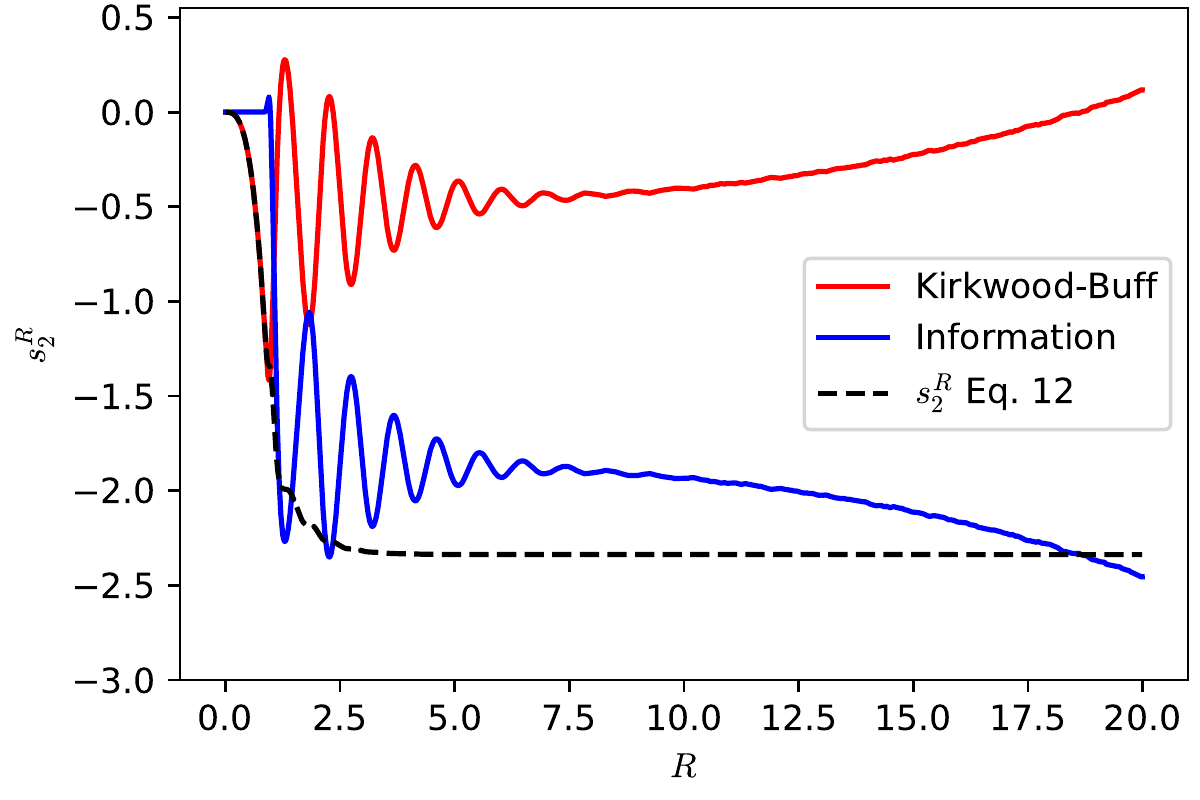}
	\caption{
Plot of the two contributions, \emph{Kirkwood-Buff} ($g(r;N_0) -1$) and \emph{Information} ($g(r;N_0)\ln g(r;N_0)$), to the truncated integral $s_2^R$ for a system of linear size $L/\sigma_{\rm LJ} = 35$ at $k_{\rm B}T=2.0\epsilon$. It is apparent that the two terms oscillate out-of-phase for small values of $R$, and their sum converges to  $s_2^{\infty}$ when $R\to \infty$.
	} \label{fig:Figure_01a}
\end{figure}
To finish this section, we compare our results with the usual truncation of Eq.~\eqref{eq:s2_TL} up to a cutoff radius $R$, namely
\begin{equation}\label{eq:s2_R}
\begin{split}
s_2^{R} = -2\pi \rho\int_{0}^{R} dr\, r^{2} &\left[ g(r;N_0)\ln g(r;N_0) \right. \\
&\left.  - (g(r;N_0) -1)  \right]\, .
\end{split}
\end{equation}
We use this truncated integral to verify numerically that ensemble finite-size contributions cancel out almost exactly.~\cite{widom2019first} For a system of size $L/\sigma_{\mathrm{LJ}} = 35$ at $k_{\rm B}T=2.0\epsilon$, we separate the $g(r;N_0) -1$, \emph{Kirkwood-Buff}, and the $g(r;N_0)\ln g(r;N_0)$, \emph{Information}, contributions and plot them as a function of the truncation radius $R$ (See Figure~\ref{fig:Figure_01a}). Both integrals diverge for large values of $R$, Kirkwood-Buff to infinity and Information to minus infinity, which signals a clear ensemble finite-size effect. However, these two finite-size contributions balance each other, and the sum of the two integrals converges to $s_2^{\infty}$ for $R>>1$. Due to this error cancellation, the truncation Eq.~\eqref{eq:s2_R} gives $s_2^{\infty}$ even for relatively small simulation boxes, and its finite-size dependence has been commonly overlooked in the literature. 
\section{Finite-volume excess entropy}\label{sec:fvs2}
Based on previous work on finite-size isothermal compressibility~\cite{Roman1999} and Kirkwood-Buff integrals~\cite{Sevilla2022}, we define a finite-volume two-body excess entropy as follows.
\begin{equation}
s_{2}(V) = -\frac{\rho}{2 V}\int \int d\mathbf{r}_1\, d\mathbf{r}_2\, R(\mathbf{r}_1)\, R(\mathbf{r}_2)\, h(\mathbf{r})\,  ,
\end{equation}
with $R(\mathbf{r})$ a step function that defines the finite integration subdomain, being equal to one inside and to zero outside the volume $V$~\cite{Roman1999}.  The function $h(\mathbf{r})$ is defined as
\begin{equation}
h(\mathbf{r})= g(\mathbf{r})\ln g(\mathbf{r}) - (g(\mathbf{r}) -1)\, .
\end{equation}
We write the double integral of $s_2(V)$ in Fourier space and include the periodicity of the simulation of the box in  $h(\mathbf{r})$ explicitly. Thus
\begin{equation}\label{eq:s2_finPBC}
s_{2}(V) = -\frac{\rho}{2 (2\pi)^3 V}\int  d\mathbf{k}\, \tilde{R}(\mathbf{k})\, \tilde{R}(-\mathbf{k})\, \tilde{h}^{\rm PBC}(\mathbf{k})\, ,
\end{equation}
where~\cite{Roman1999}
\begin{equation}\label{eq:hPBC}
\tilde{h}^{\rm PBC}(\mathbf{k})=\sum_{n_x,n_y,n_z} e^{-\mathbf{k}\cdot\mathbf{s}_{n_x,n_y,n_z}} \tilde{h}(\mathbf{k})\, ,
\end{equation}
with $\tilde{h}(\mathbf{k})$ the Fourier transform of $h(\mathbf{r})$ and $\mathbf{s}_{n_x,n_y,n_z}=(n_x\, L_{x},n_y\, L_{y},n_z\, L_{x})$ a vector specifying the system's periodic images such that $n_{x,y,z}$ takes integer values. In the following, we consider a cubic simulation box with $L_{x}=L_{y}=L_{z}=L$. As before~\cite{Sevilla2022}, we choose $|n_x|\le 1$, $|n_y|\le 1$ and  $|n_z|\le 1$ to compute Eq.~\eqref{eq:hPBC}. Finally, we assume a homogeneous and isotropic fluid such that $\tilde{h}(\mathbf{k})=\tilde{h}(k)$ with $k=\sqrt{\mathbf{k} \cdot \mathbf{k} }$.\\
\begin{figure}[h]
	\centering
	\includegraphics[width=0.48\textwidth]{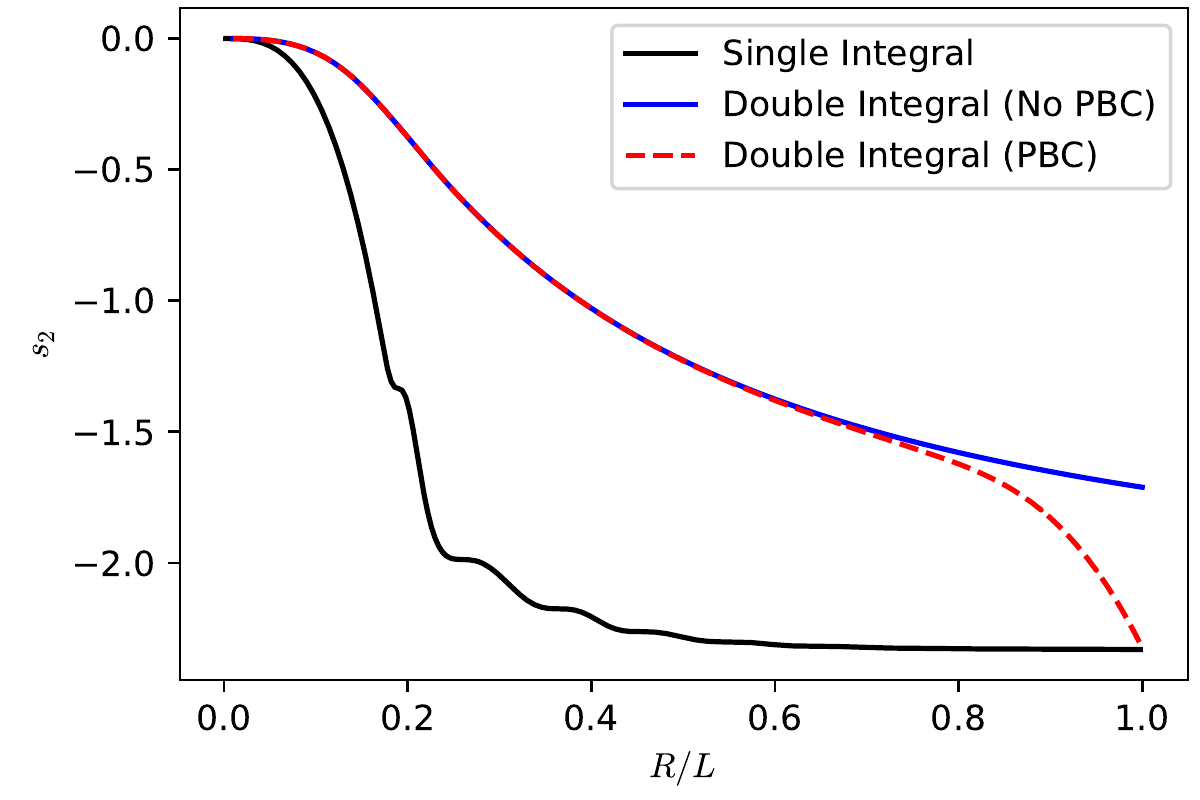}
	\caption{Running $s_2$ as a function of the ratio $R/L$ for the case $L/\sigma_{\mathrm{LJ}}=5$ at $k_{\rm B}T=2.0\epsilon$. The black line corresponds to the truncation Eq.~\ref{eq:s2_R}, and the red and blue curves are the result of Eq.~\eqref{eq:s2_finPBC} including ($|n_x|\le 1$, $|n_y|\le 1$ and  $|n_z|\le 1$) and not including ($|n_x|=|n_y|=|n_z|= 0$) PBC, respectively. By including PBC, the integral Eq.~\eqref{eq:s2_finPBC} converges to the thermodynamic limit.}
	\label{fig:Figure_02}
\end{figure}
To validate our approach, we verify that Eqs~\eqref{eq:s2_finPBC} and \eqref{eq:s2_R} converge to the same value in the thermodynamic limit. %\robin{Mauricio: perhaps we need to consider here the largest system we have.}
To  this aim, we consider a system with linear size $L/\sigma_{\mathrm{LJ}} = 50$ at $k_{\rm B}T=2.0\epsilon$, compute the RDF and evaluate the truncated integral Eq.~\eqref{eq:s2_R}. According to Eq.~\eqref{eq:s2_1overL}, implicit  finite-size effects are the most relevant in this case. Hence, by considering a sufficiently large simulation box, the large $R$ limit of Eq.~\eqref{eq:s2_R} converges to the TL value. We present this result in Fig.~\ref{fig:Figure_02} (black solid curve).  To evaluate Eq.~\eqref{eq:s2_finPBC}, we take the RDF from the simulation box with linear size $L/\sigma_{\mathrm{LJ}}=20$ and perform the Fourier transform procedure described above to obtain $\tilde{h}(\mathbf{k})$. It is apparent, as expected, that with explicit PBC, the finite-size $s_2$ gives the TL value (red dashed curve). Instead, by removing PBC, there is a significant deviation from the TL value that we attribute to the $1/L$ dependence in Eq.~\eqref{eq:s2_1overL} (blue solid curved). \\
\begin{figure}[h]
	\centering
	\includegraphics[width=0.48\textwidth]{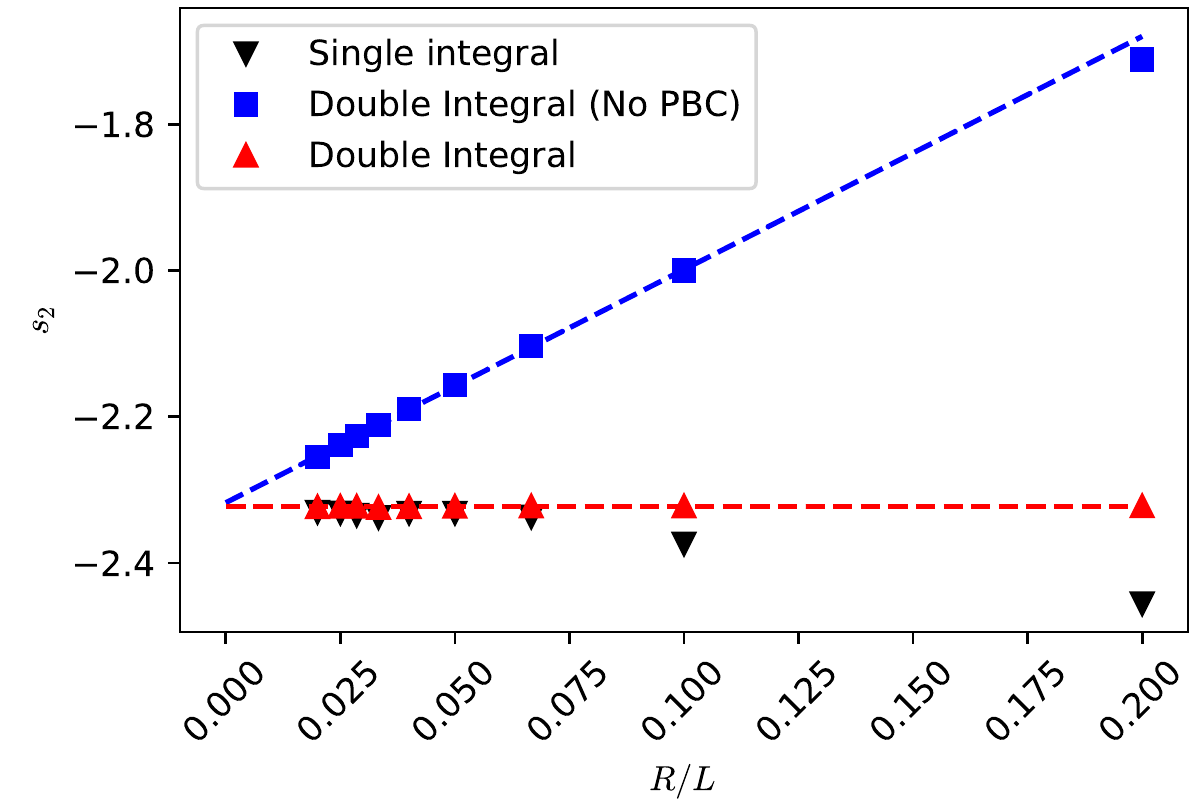}
	\caption{$s_2$ as a function of the inverse of the simulation box size $L$ for systems at $k_{\rm B}T=2.0\epsilon$. The black triangles are calculated with the truncated integral (Eq.~\eqref{eq:s2_R}), the red triangles and blue squares were calculated with the double integral (Eq.~\eqref{eq:s2_finPBC}) including and excluding PBC, respectively.}
	\label{fig:Figure_03}
\end{figure}
We verify this $1/L$ dependence in the finite-size $s_2$. In Figure~\ref{fig:Figure_03}, we plot  the result of $s_2^{R}$, Eq.~\eqref{eq:s2_R}, as a function of $1/L$ (black inverted triangles). There, it is apparent that the integral converges when the linear size of the system is $L/\sigma_{\mathrm{LJ}}>10$. The result of using $s_2(V)$, Eq.~\eqref{eq:s2_finPBC}, with explicit PBC, always converges to the TL value (red triangles), regardless of the linear size of the system. More interestingly, by removing PBC from Eq.~\eqref{eq:s2_finPBC}, we observe a clear linear dependence with $1/L$ (blue squares). Furthermore, by extrapolating this behaviour (blue dashed line) to the axis $1/L=0$, we obtain a linear extrapolation to $s_2^{\infty}$. This result completes the validation of both, Eqs~\eqref{eq:s2_1overL} and \eqref{eq:s2_finPBC}. 
\section{Finite-size excess-entropy scaling}\label{sec:fsDs2}
\begin{figure}[!ht]
	\centering
	\includegraphics[width=0.48\textwidth]{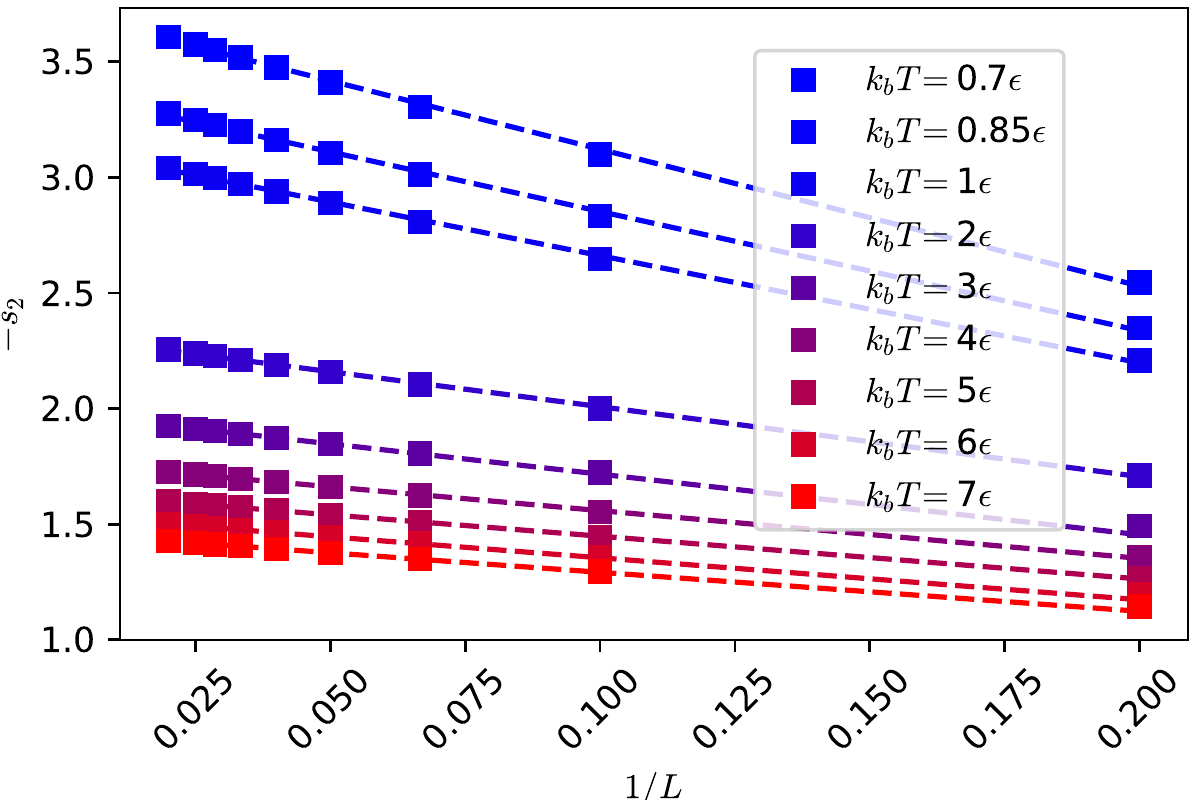}
	\caption{-$s_2$ as a function of $1/L$ for a LJ system at $\rho\sigma_{\mathrm{LJ}}^3=0.864$ and different temperatures. All data points were obtained with the RDF for the system of linear size $L/\sigma_{\mathrm{LJ}}=20$ and using Eq.~\eqref{eq:s2_finPBC} without PBC.}
	\label{fig:Figure_04}
\end{figure}
\begin{figure}[h]
	\centering
	\includegraphics[width=0.48\textwidth]{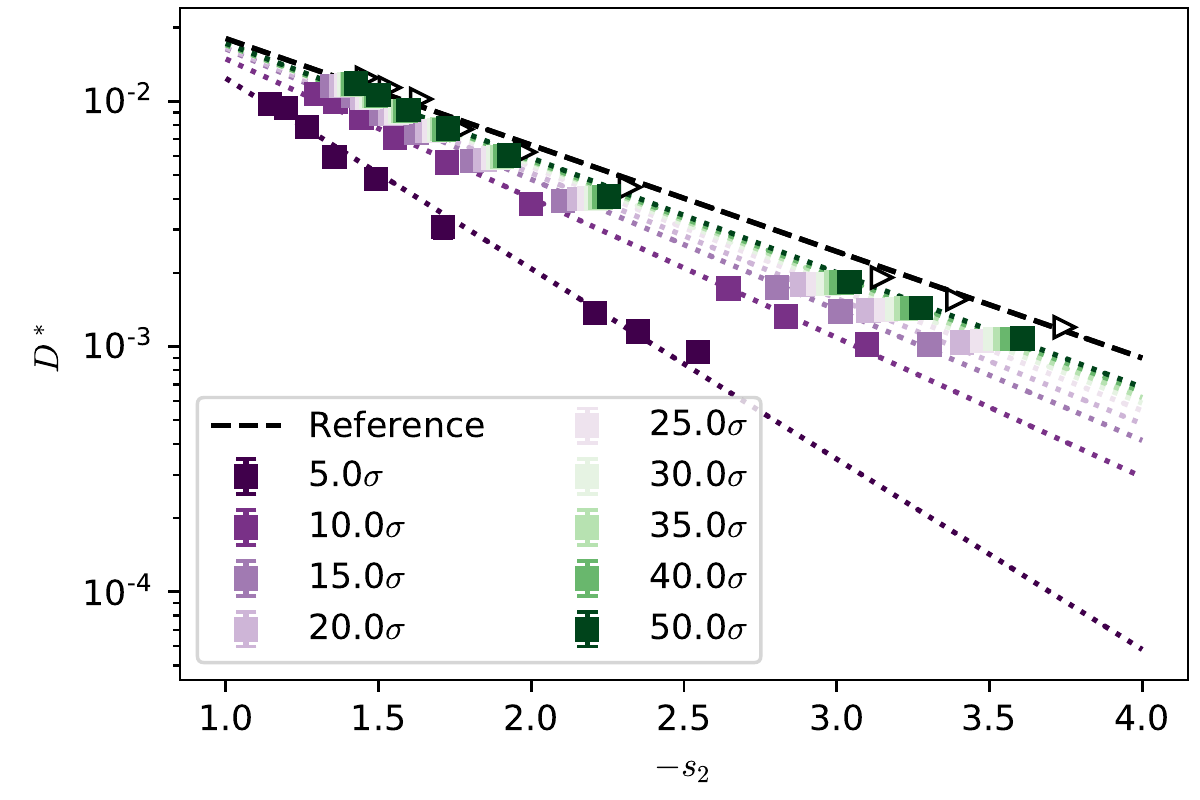}
	\caption{Reduced self-diffusion coefficient $D^*$ as a function of $-s_2$ for different system sizes and temperatures. The empty black triangles show the thermodynamic limit values for $-s_2$ and $D^*$.}
	\label{fig:Figure_05}
\end{figure}
\begin{figure}[h]
	\centering
	\includegraphics[width=0.48\textwidth]{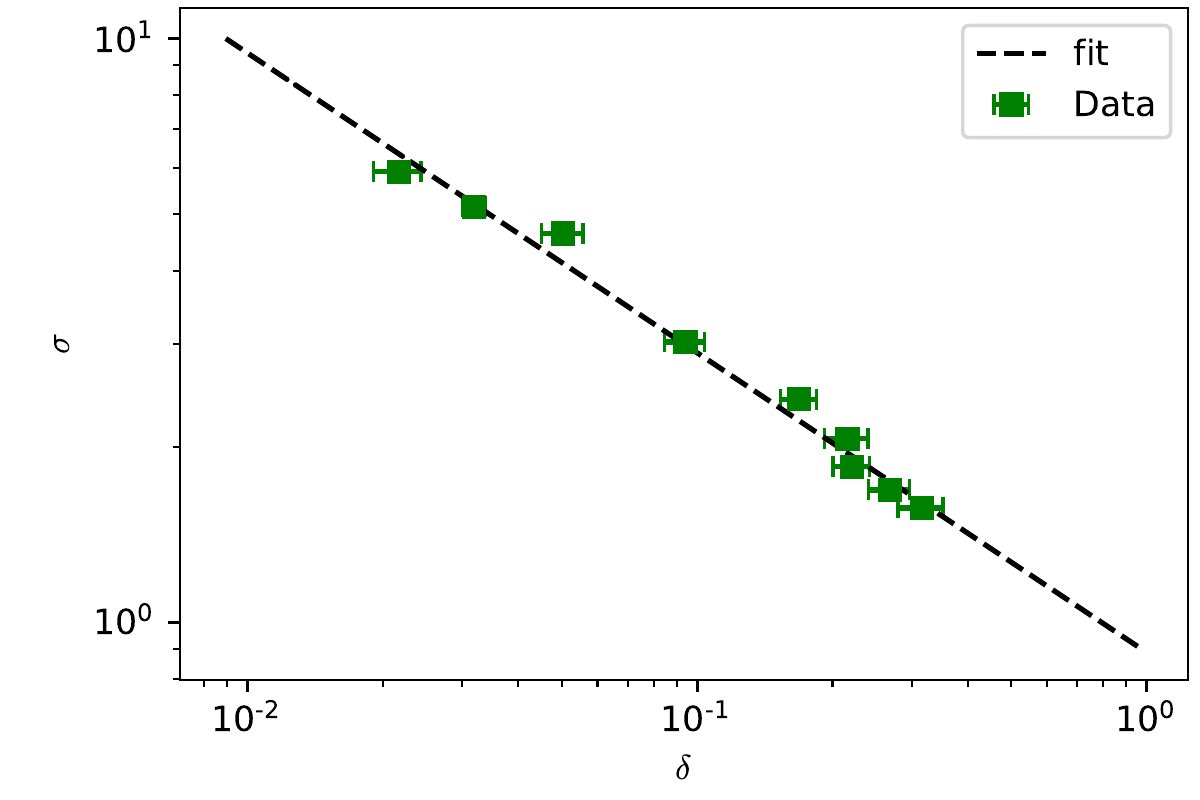}
	\caption{Coefficients $\sigma(T)$ as a function of $\delta(T)$ for all the temperatures considered here. We used a power law $\sigma=a\delta^b$ to fit the data. }
	\label{fig:Figure_06}
\end{figure}

In this section, we investigate the finite-size effects of the self-diffusivity entropy scaling, Eq.~\eqref{eq:Dzugutov}. To this aim, we verify that the scaling of $s_2$ with $1/L$ is valid in a wide temperature range. We present these results for a LJ system with density $\rho\sigma_{\mathrm{LJ}}^3=0.864$ in the range of temperatures $k_{\rm B}T=[0.7\epsilon,7\epsilon]$. The results in Figure \ref{fig:Figure_04} indicate that the  $1/L$ scaling is apparent for all temperatures considered here.\\
We now collect all our data to investigate the scaling of Eq.~\eqref{eq:Dzugutov} with the simulation box size. The result is presented in Figure \ref{fig:Figure_05} where the diffusion constant $D^*$ is plotted against $-s_2$. A clear trend with system size emerges, indicating that Eq.~\eqref{eq:Dzugutov} remains valid even for the smallest simulation boxes considered and showing that the parameters $A$ and $\alpha$ are also size dependent. By extrapolating $D^*$ and $-s_2$ to the limit $1/L\to 0$, we obtain the TL values given by the black empty triangles that well agree with the reference scaling provided by Eq.~\eqref{eq:Dzugutov} (black dashed line). Indeed, we report $A^\infty=0.048\pm 0.001$ and $\alpha^\infty=1.000\pm 0.013$ in the TL, in good agreement with the value originally estimated in Ref.~\cite{Dzugutov1996}.\\
Finally, we investigate the relation between the coefficients $\delta$ and $\sigma$ of the finite-size scaling of $D^*$ and $s^2$, respectively. In Figure \ref{fig:Figure_06}, we plot $\sigma$ as a function of $\delta$ and observe a power law relation of the form $\sigma = a\delta^b$ with $a=1.256 \pm 0.118$ and $b=-0.513\pm 0.020$. 
\section{Summary and outlook}\label{sec:Con}
We define a finite-size two-body excess entropy $s_2(L)$ integral equation with $L$ the linear size of the simulation box. Using analytical arguments and simulations of a prototypical Lennard-Jones liquid at different densities and temperatures, we show that $s_2(L)=s_2^{\infty}+\sigma/L$ with $\sigma$ a constant that depends on intensive thermodynamic quantities. Given the well-know finite-size scaling of the self-diffusivity, $D^*(L)=D^{*\infty}-\delta/L$, we show that the universal scaling relation between entropy and diffusion $D^{*}=A\exp{(\alpha s_2)}$ also exhibits a finite-size dependence and, by extrapolating to the TL, report $A=0.048\pm 0.001$ and $\alpha=1.000\pm 0.013$, in good agreement with values reported in the literature. Finally, and perhaps more interestingly,  we show that the scaling coefficients $\sigma$ and $\delta$ of $s_2$ and $D^*$, respectively, are related by a somewhat simple power law $\sigma = a\delta^b$ with $a=1.256\pm 0.118$ and $b=-0.513\pm 0.020$. \\
The finite-size scaling of $s_2$ can be rationalised in terms of the thermodynamics of small systems~\cite{hill,Puglisi_etal_Entropy2018}. In particular, the statistical mechanics of a few model small systems in confinement has been derived recently~\cite{Braten_etal_JCP2021}. The authors have shown that given the high surface area--to--volume ratio of small systems, thermodynamic properties include surface contributions. In the case of entropy, these contributions include $1/L$ terms with $L$, the linear size of the system. In this context, we feel that the finite-size entropy scaling investigated here might play a role in understanding the non-equilibrium thermodynamics of confined, small systems~\cite{nanothermobook}.\\
The power law relation between the scaling coefficients of self-diffusion and two-body excess entropy is somewhat intriguing.  On the one hand, the size scaling in the self-diffusion appears as a consequence of the conservation of linear momentum~\cite{Hummer2004}. On the other hand, the finite-size scaling in the two-body entropy results from a surface contribution due to the confinement of the system~\cite{Braten_etal_JCP2021}. Admittedly, we do not have a satisfactory explanation for this connection.\\ 
 Nevertheless, we point out that the ratio $\delta^{b}/\sigma=1/a$ might be related to a constant viscosity/entropy ratio. Indeed, $\delta$ is inversely proportional to the system's viscosity, and a simple dimensional analysis tells us that $\sigma$ has units of entropy times length.  Interestingly, string theory methods have been used to conjecture that, for fluids in equilibrium, the viscosity to entropy density ratio has a lower bound at $\hbar/4\pi k_{\rm B}$~\cite{Kovtun_etal_PRL2005} with $\hbar$ the reduced Planck constant. This relation, tested for various fluid systems~\cite{Angiella_etal_PhysLettA2009, Faussurier_etal_HEDP2014, Hohm_ChemPhys2014}, has been originally derived by considering that the entropy density of a black hole is proportional to the surface to volume ratio of its event horizon, i.e. a $1/L$ contribution. We find this connection fascinating, and, in our opinion, it deserves further investigation.\\
\acknowledgments
We are grateful to Kurt Kremer for his insightful discussions. We also thank Denis Andrienko for his critical reading of the manuscript. R.C.-H. gratefully acknowledges funding from SFB-TRR146 of the German Research Foundation (DFG). Simulations have been performed on the THINC cluster at the Max Planck Institute for Polymer Research and the COBRA cluster at the Max Planck Computing and Data Facility. 
%
%    
%\nocite{*}
%\bibliographystyle{unsrt}%plain
%\bibliography{references.bib}
%

%
\end{document}